\documentclass[aps,superscriptaddress,amsmath,amssymb,twocolumn,floatfix,english]{revtex4}

\usepackage{url}
\usepackage{bm}
\usepackage{graphicx}
\usepackage{amsmath}
\usepackage{amstext}
\usepackage{amssymb}
\usepackage{amsfonts}
\usepackage{amsbsy}
\usepackage{verbatim}
\usepackage{color}
\usepackage[colorlinks=true, urlcolor=blue, linkcolor=blue, citecolor=blue, pdftex]{hyperref}
\usepackage{multirow}
\usepackage{pdfpages}
\usepackage{floatrow}
\usepackage{gensymb}
\usepackage{textcomp}

\newcommand{\dagga}{{\phantom{\dagger}}}

\begin{document}

\title{Magnetic and spin-liquid phases in the frustrated $t-t^\prime$ Hubbard model on the triangular lattice}

\author{Luca F. Tocchio}
\affiliation{Institute for Condensed Matter Physics and Complex Systems, DISAT, Politecnico di Torino, I-10129 Torino, Italy}
\author{Arianna Montorsi}
\affiliation{Institute for Condensed Matter Physics and Complex Systems, DISAT, Politecnico di Torino, I-10129 Torino, Italy}
\author{Federico Becca}
\affiliation{Dipartimento di Fisica, Universit\`a di Trieste, Strada Costiera 11, I-34151 Trieste, Italy}

\date{\today}

\begin{abstract}
The Hubbard model and its strong-coupling version, the Heisenberg one, have been widely studied on the triangular lattice to capture the 
essential low-temperature properties of different materials. One example is given by transition metal dichalcogenides, as 1T$-$TaS$_2$,
where a large unit cell with $13$ Ta atom forms weakly-coupled layers with an isotropic triangular lattice. By using accurate variational 
Monte Carlo calculations, we report the phase diagram of the $t-t^\prime$ Hubbard model on the triangular lattice, highlighting the 
differences between positive and negative values of $t^\prime/t$; this result can be captured only by including the charge fluctuations 
that are always present for a finite electron-electron repulsion. Two spin-liquid regions are detected: one for $t^\prime/t<0$, which 
persists down to intermediate values of the electron-electron repulsion, and a narrower one for $t^\prime/t>0$. The spin-liquid phase 
appears to be gapless, though the variational wave function has a nematic character, in contrast to the Heisenberg limit. We do not find 
any evidence for non-magnetic Mott phases in the proximity of the metal-insulator transition, at variance with the predictions (mainly 
based upon strong-coupling expansions in $t/U$) that suggest the existence of a weak-Mott phase that intrudes between the metal and the 
magnetically ordered insulator.
\end{abstract}

\maketitle

\section{Introduction}\label{sec:intro}
 
Searching and understanding quantum spin-liquid phases is one of the key topics in contemporary condensed-matter physics~\cite{balents2010}. 
Such states are favored by the presence of frustration, being realized in lattices with competing magnetic interactions. In particular, 
strong evidences that support the presence of a spin liquid are reported in Herbertsmithite, well described by the Heisenberg model on the 
kagome lattice~\cite{norman2016}, and for organic compounds like $\kappa$(ET)$_2$Cu$_2$(CN)$_3$ and Me$_3$EtSb[Pd(dmit)$_2$]$_2$, whose 
low-temperature behavior could be captured by the Hubbard model on the anisotropic triangular lattice~\cite{kanoda2011,powell2011}.
Recently, a transition metal dichalcogenide, 1T-TaS$_2$, came to the attention of the community working on spin liquids~\cite{law2017}.
Indeed, this compound was observed to undergo a low-temperature transition into a cluster of stars of David, where the unit cell contains 
13 Ta atoms and form an isotropic triangular lattice. The low-temperature behavior is compatible with a pure Mott insulator, with no
long-range magnetic order~\cite{wilson1975,disalvo1977,fazekas1979}. Still, charge fluctuations are present and the material is expected 
to be not too far from a metal-insulator transition. In the past, the issue of magnetism has not been discussed much in the literature, 
while recent NMR and $\mu$SR experiments highlighted the absence of static magnetic moments~\cite{klanjsek2017,kratochvilova2017}.
This information, together with indications from NMR of a weak inter-layer coupling, suggests that the system may be a good candidate for 
hosting a spin-liquid phase. 

The theoretical investigation of spin-liquid phases on isotropic triangular lattices has been mostly confined to spin $S=1/2$ models, where 
spin liquids can be systematically classified, according to the projective symmetry group theory~\cite{wang2006,lu2015}, also including 
the effect of gauge fluctuations~\cite{song2019,song2020}. Starting from the Heisenberg model with nearest-neighbor (NN) super-exchange $J$,
spin-liquid phases can be stabilized by including either a next-nearest-neighbor (NNN) coupling $J^\prime$ or a four-spin ring-exchange term 
$K$. The latter one can be justified within the fourth-order strong-coupling expansion in $t/U$ and is usually considered for an effective
description of density fluctuations close to the Mott transition~\cite{grover2010}. The case with $J^\prime$ has been widely investigated:
In the classical limit, there is a three-sublattice order for $J^\prime/J < 1/8$, where each spin is oriented with a 120$^{\circ}$ angle 
with respect to its nearest neighbors; for $1/8<J^\prime/J<1$, the lowest-energy state is highly degenerate, with configurations having spins 
summing to zero on each 4-site rhomboidal plaquette; for larger values of $J^\prime/J$, spiral states are obtained. When quantum fluctuations 
are included (e.g., within the spin-wave approximation), a paramagnetic phase emerges in the proximity of the classical transition 
$J^\prime/J=1/8$; in addition, quantum corrections give rise to an order-by-disorder selection for $1/8 \lesssim J^\prime/J \lesssim 1$, 
leading to a stripe collinear order with 4 out of 6 nearest-neighbor correlations being antiferromagnetic and the remaining 2 being 
ferromagnetic~\cite{jolicoeur1990,chubukov1992,chitra1995}.
Recently, this model has been analysed by using variational Monte Carlo (VMC) and density-matrix renormalization group (DMRG) approaches. 
In the former case, a gapless spin liquid has been first proposed in Ref.~\onlinecite{kaneko2014} and later confirmed~\cite{iqbal2016}. 
Within this scenario, the ground state could be well approximated by a fermionic Gutzwiller-projected wave function, having Dirac points 
in the spinon band and emergent U(1) gauge fields. Within the DMRG approach, some initial calculations suggested the presence of a gapped 
spin liquid~\cite{zhu2015,hu2015}, while more recent ones also pointed towards the possibility of a gapless spin liquid~\cite{hu2019}. 
Furthermore, in the presence of ring-exchange terms $K$, a gapless spin liquid with a Fermi surface has been proposed by earlier VMC 
studies~\cite{motrunich2005}, as well by recent DMRG ones~\cite{he2018}, for large enough values of the ratio $K/J$. Another VMC study 
proposed instead two possible spin liquids, as a function of $K/J$: a gapless nodal $d$-wave one and another one with a quadratic band 
touching, both without a spinon Fermi surface~\cite{mishmash2013}.  

The hunt for spin liquids in the presence of change fluctuations, i.e., within the Hubbard model, is instead more limited. Indeed, early 
Hartree-Fock calculations~\cite{krishnamurty1990,jayaprakash1991} concentrated the attention on the structure of the magnetic order across 
the Mott transition. Since then, different approaches have been applied to understand whether a spin liquid phase can be stabilized close 
to the Mott transition (the so-called weak-Mott insulator), between the metal-insulator transition and the insurgence of magnetic order. 
The outcomes are not conclusive: calculations based upon variational cluster approximation (VCA)~\cite{sahebsara2008,yamada2014,laubach2015}, 
path-integral renormalization group~\cite{yoshioka2009}, strong-coupling expansion~\cite{yang2010}, dual-fermion approach~\cite{antipov2011}, 
and DMRG~\cite{shirakawa2017,szasz2018} suggested the existence of an intermediate spin-liquid phase; by contrast, a direct transition between 
a metal and a magnetic insulator has been found by using dynamical cluster approximation~\cite{lee2008} and VMC~\cite{watanabe2008,tocchio2013}. 
This analysis is complicated by the significant difference in locating the Mott transition observed with the different methods. Recently, a 
calculation of magnetic and charge susceptibilities has been attempted, which, however, could not reach sufficiently low temperatures to assess 
the existence of a spin-liquid phase~\cite{li2020}. The effect of next-nearest neighbor hopping has been addressed in Ref.~\onlinecite{misumi2017}, 
using the VCA method with few ($12$) sites, leading to a large spin-liquid region for $t^\prime/t>0$.

In this paper, we consider the Hubbard model on a triangular lattice with both NN and NNN hoppings, in order to increase the role of magnetic 
frustration, thus favoring spin-liquid phases. We employ variational wave functions and Monte Carlo sampling to evaluate ground-state properties
and draw the phase diagram in the $(t^\prime/t,U/t)$ plane. The main outcome is that the stability of the spin-liquid phase depends both on the 
degree of frustration, i.e., $(t^\prime/t)^2=J^\prime/J$, and on the Fermi surface topology at small values of $U/t$. This combination of strong-
and weak-coupling physics is crucial in understanding how stable a spin-liquid phase is when charge fluctuations are taken into account. 
In particular, when the ratio $t^\prime/t$ falls within the spin-liquid regime of the Heisenberg model, the case with $t^\prime/t<0$ hosts 
a spin liquid down to intermediate values of $U/t$, where the stripe collinear order becomes competitive, while the case with $t^\prime/t>0$
is dominated by the coplanar 120$^\circ$ order. The spin liquid in the Hubbard model appears to be nematic and presumably gapless. We remark 
that we do not find any evidence for a weak-Mott insulator, thus posing doubts on the validity of strong-coupling expansions down to the Mott
transition.

\section{Model and method}\label{sec:method}

We consider the single-band Hubbard model on the triangular lattice:
\begin{equation}\label{eq:hubbard}
\begin{split}
{\cal H} = & -t \sum_{\langle i,j\rangle,\sigma} c^\dagger_{i,\sigma} c_{j,\sigma}^{\phantom{\dagger}} 
 - t^\prime \sum_{\langle\langle i,j\rangle\rangle,\sigma} c^\dagger_{i,\sigma} c_{j,\sigma}^{\phantom{\dagger}} + \textrm{H.c.} \\
           & +U \sum_{i} n_{i,\uparrow} n_{i,\downarrow}\,,
\end{split}
\end{equation}
where $c^\dagger_{i,\sigma}$ ($c^\dagga_{i,\sigma}$) creates (destroys) an electron with spin $\sigma$ on site $i$ and 
$n_{i,\sigma}=c^\dagger_{i,\sigma} c^\dagga_{i,\sigma}$ is the electronic density per spin $\sigma$ on site $i$. The NN and NNN hoppings 
are denoted as $t$ and $t^\prime$, respectively; $U$ is the on-site Coulomb interaction. We define three vectors connecting NN sites, 
${\bf a}_1=(1,0)$, ${\bf a}_2=(1/2,\sqrt{3}/2)$, and ${\bf a}_3=(-1/2,\sqrt{3}/2)$; in addition, we also define three vectors for NNN sites, 
${\bf b}_1={\bf a}_1+{\bf a}_2$, ${\bf b}_2={\bf a}_2+{\bf a}_3$, and ${\bf b}_3={\bf a}_3+{\bf a}_1$. In the following, we consider clusters 
with periodic boundary conditions defined by ${\bf T}_1=l {\bf a}_1$ and ${\bf T}_2=l {\bf a}_2$, in order to have $l \times l$ lattices with 
$L=l^2$ sites. The half-filled case, which is relevant for the spin-liquid physics, is considered here. In this case, only the sign of the
ratio $t^\prime/t$ is relevant and not the individual signs of $t$ and $t^\prime$.

Our numerical results are obtained by means of the VMC method, which is based on the definition of suitable wave functions to approximate the
ground-state properties beyond perturbative approaches~\cite{becca2017}. In particular, we consider the so-called Jastrow-Slater wave functions
that include long-range electron-electron correlations via the Jastrow factor~\cite{capello2005,capello2006}, on top of an uncorrelated Slater
determinant (possibly including electron pairing). In  addition, the so-called backflow correlations will be applied to the Slater determinant,
in order to sizably improve the quality of the variational state~\cite{tocchio2008,tocchio2011}. Thanks to Jastrow and backflow terms, these
wave functions can reach a very high degree of accuracy in Hubbard-like models, for different regimes of parameters, including frustrated 
cases~\cite{leblanc2015}. Therefore, they represent a valid tool to investigate strongly-correlated systems, competing with state-of-the-art 
numerical methods, as DMRG or tensor networks.

Our variational wave function for describing the spin-liquid phase is defined as:
\begin{equation}\label{eq:wf_BCS}
|\Psi_{\textrm{BCS}}\rangle={\cal J}_d |\Phi_{\textrm{BCS}}\rangle\,,
\end{equation}
where ${\cal J}_d$ is the density-density Jastrow factor and $|\Phi_{\textrm{BCS}}\rangle$ is a state where the orbitals of an auxiliary 
Hamiltonian are redefined on the basis of the many-body electronic configuration, incorporating virtual hopping processes, via the backflow 
correlations~~\cite{tocchio2008,tocchio2011}. The auxiliary Hamiltonian for the spin-liquid wave function is defined as follows:
\begin{equation}\label{eq:H_BCS}
{\cal H}_{\textrm{BCS}} = \sum_{k,\sigma} \xi_k c^{\dagger}_{k,\sigma}c^{\dagga}_{k,\sigma}
+\sum_{k}\Delta_k c^{\dagger}_{k,\uparrow}c^{\dagger}_{-k,\downarrow}+\textrm{H.c.}\,,
\end{equation}
where $\xi_k=\tilde{\epsilon}_k-\mu$ defines the free-band dispersion (including the chemical potential $\mu$) and $\Delta_k$ is the singlet 
pairing amplitude. By performing a particle-hole transformation on spin-down electrons, the Hamiltonian commutes with the particle number and,
therefore, ``orbitals'' may be defined (with both spin-up and spin-down components). In the Heisenberg model, different choices for $\xi_k$ 
and $\Delta_k$ lead to distinct spin liquids, which have been systematically classified~\cite{lu2015}. This classification is not any more 
rigorous in the Hubbard model; indeed, most of them cannot be stabilized for finite values of $U/t$. Instead, we find that the best spin-liquid 
is characterized by anisotropic parameters in the auxiliary Hamiltonian. The hopping terms are given by:
\begin{eqnarray}
&& \tilde{\epsilon}_k = -2t \left[ \cos({\bf k} \cdot {\bf a}_2) + \cos({\bf k} \cdot {\bf a}_3) \right] \nonumber \\
&& -2 \tilde{t}^\prime \left[ \cos({\bf k} \cdot {\bf b}_1) + \cos({\bf k} \cdot {\bf b}_2) + \cos({\bf k} \cdot {\bf b}_3) \right ].
\label{eq:hopp}
\end{eqnarray}
Instead, the pairing amplitudes are:
\begin{equation}\label{eq:pair}
\Delta_k=2\Delta_{\textrm{BCS}} \left[ \cos({\bf k} \cdot {\bf a}_2) -\cos({\bf k} \cdot {\bf a}_3) \right],
\end{equation}
which possess a $d$-wave symmetry on the two bonds with finite variational hoppings. The broken rotational symmetry in $\tilde{\epsilon}_k$ and $\Delta_k$ 
will naturally lead to nematicity (e.g., different spin-spin correlations along NN bonds). Since the variational state has no magnetic order, it 
describes a nematic $d$-wave spin liquid. This {\it Ansatz} has been compared with the U(1) Dirac spin liquid that has been suggested by the VMC 
study of the Heisenberg model with NN and NNN couplings of Ref.~\onlinecite{iqbal2016}. However, such state has a poor energy for finite values 
of $U/t$ (at least, up to $U/t\sim 25$). We have also tested the following two states with the symmetries of the triangular lattice: 
i) A $\mathbb{Z}_2$ state with uniform hoppings and pairings at NN and NNN amplitudes and ii) a complex-pairing state, with uniform hopping along 
NN and NNN bonds and a pairing $\Delta_k=2\Delta_{\textrm{BCS}} \left[ \cos({\bf k} \cdot {\bf a}_1)+\omega \cos({\bf k} \cdot {\bf a}_2) + 
\omega^2 \cos({\bf k} \cdot {\bf a}_3)\right]$, where $\omega=e^{2i\pi/3}$. While the state i) is not stable upon optimization, the state ii) 
can be stabilized, but with an energy higher than the nematic state. Finally, we have checked that chiral states can be also stabilized, but 
with an energy higher than the $d$-wave optimal state. In this respect, we have considered both complex hoppings in the auxiliary Hamiltonian 
of Eq.~(\ref{eq:H_BCS}), as discussed in Ref.~\onlinecite{hu2016} for the Heisenberg model, and the so-called $d+id$ pairing symmetry with 
$\Delta_k=2\Delta_1 \left[ \cos({\bf k} \cdot {\bf a}_2) -\cos({\bf k} \cdot {\bf a}_3) \right]+2i\Delta_2 \cos({\bf k} \cdot {\bf a}_1)$.

The density-density Jastrow factor is ${\cal J}_d = \exp \left ( -1/2 \sum_{i,j} v_{i,j} n_{i} n_{j} \right )$, where 
$n_{i}= \sum_{\sigma} n_{i,\sigma}$ is the electron density on site $i$ and $v_{i,j}$ are pseudopotentials that are optimized for every 
independent distance $|{\bf R}_i-{\bf R}_j|$. The density-density Jastrow factor allows us to describe a nonmagnetic Mott insulator for a 
sufficiently singular Jastrow factor $v_q \sim 1/q^2$ ($v_q$ being the Fourier transform of $v_{i,j}$)~\cite{capello2005,capello2006}.

Our variational wave function for the magnetic phases is defined as:
\begin{equation}
|\Psi_{\textrm{AF}}\rangle={\cal J}_s{\cal J}_d |\Phi_{\textrm{AF}}\rangle\,,
\end{equation}
where ${\cal J}_s$ is the spin-spin Jastrow factor and $|\Phi_{\textrm{AF}}\rangle$ is obtained, after taking into account the backflow 
corrections, from the following auxiliary Hamiltonian:
\begin{equation}\label{eq:AF}
{\cal H}_{\textrm{AF}}=\sum_{k,\sigma}\epsilon_k c^{\dagger}_{k,\sigma}c^{\dagga}_{k,\sigma}+\Delta_{\textrm{AF}}\sum_{i}{\bf M}_i\cdot{\bf S}_i,
\end{equation}
where $\epsilon_k$ is the free dispersion of Eq.~(\ref{eq:hubbard}), ${\bf S}_i$ is the spin operator at site $i$ and ${\bf M}_i$ is defined as 
${\bf M}_i=[\cos({\bf Q}\cdot {\bf R}_i),\sin({\bf Q}\cdot {\bf R}_i),0]$, where ${\bf Q}$ is the pitch vector. The three-sublattice 120$^{\circ}$ 
order has ${\bf Q}=(\frac{4\pi}{3},0)$ or $(\frac{2\pi}{3},\frac{2\pi}{\sqrt{3}})$, while the stripe collinear order with a two-sublattice 
periodicity has ${\bf Q}=(0,\frac{2\pi}{\sqrt{3}})$. Similarly to the case of density-density correlations, the spin-spin Jastrow factor is 
written in terms of a pseudopotential $u_{i,j}$ that couples the $z$-component of the spin operators on different sites. The spin-spin Jastrow 
factor describes the relevant quantum fluctuations around the classical spin state, which is defined in the $x-y$ plane~\cite{becca2000}. 

All the pseudopotentials in the Jastrow factors, the parameters $\Delta_{\textrm{BCS}}$, $\Delta_{\textrm{AF}}$, $\tilde{t}^\prime$ and $\mu$, 
as well as the backflow corrections are simultaneously optimized, while $t$ is kept fixed to 1 to set the energy scale.

In order to assess the metallic or insulating nature of the ground state we can compute the static density-density structure factor:
\begin{equation}\label{eq:Nq}
N({\bf q})=\frac{1}{L}\sum_{i,j}\langle n_in_j\rangle^{i {\bf q}\cdot({\bf R}_i-{\bf R}_j)},
\end{equation}
where $\langle\dots\rangle$ indicates the expectation value over the variational wave function. Indeed, charge excitations are gapless when 
$N({\bf q})\propto |{\bf q}|$ for $|{\bf q}| \to 0$, while a charge gap is present whenever $ N({\bf q})\propto |{\bf q}|^2$ for 
$|{\bf q}|\to 0$~\cite{feynman1954,tocchio2011}. Analogously, the presence of a spin gap can be checked by looking at the small-$q$ behavior 
of the static spin-spin correlations~\cite{tocchio2019}:
\begin{equation}\label{eq:Sq}
S({\bf q})=\frac{1}{L}\sum_{i,j}\langle S^z_i S^z_j\rangle^{i {\bf q}\cdot({\bf R}_i-{\bf R}_j)}.
\end{equation}

\begin{figure}
\includegraphics[width=0.9\textwidth]{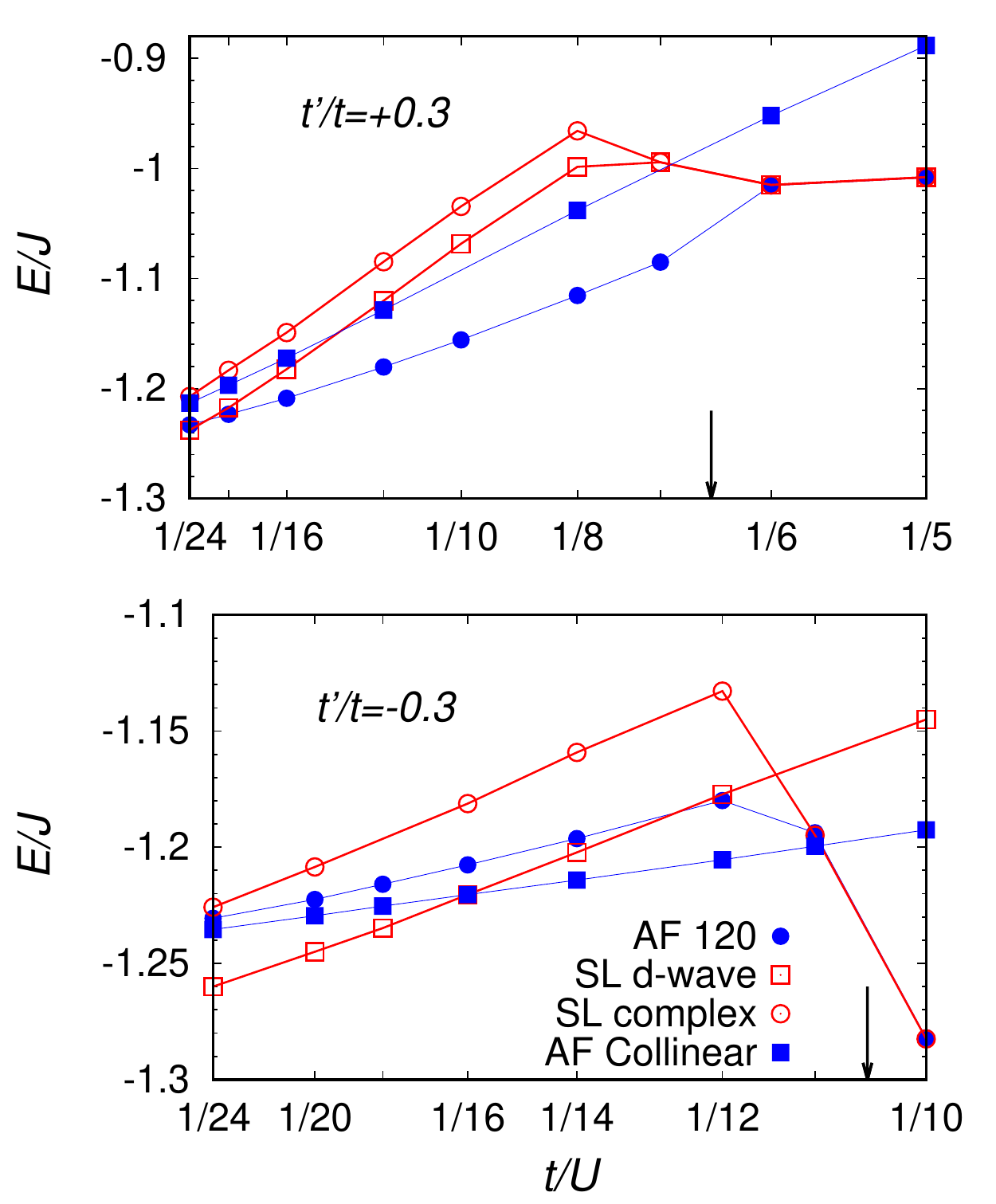}
\caption{Energy (per site) in units of $J=4t^2/U$, as a function of $t/U$ for $t^\prime/t=+0.3$ (upper panel) and $t^\prime/t=-0.3$ (lower 
panel). Data are shown for four different trial wave functions: The spin liquids ``SL $d$-wave'' (red empty squares), and ``SL complex'' 
(red empty circles), the magnetic state with the three-sublattice 120$^{\circ}$ order (blue circles), and the magnetic state with the stripe 
collinear order (blue squares). Black arrows denote the metal-insulator transitions. Data are shown for a $L=18\times 18$ lattice size. Error 
bars are smaller than the symbol size.}
\label{fig:energy}
\end{figure}

\begin{figure}
\includegraphics[width=0.9\textwidth]{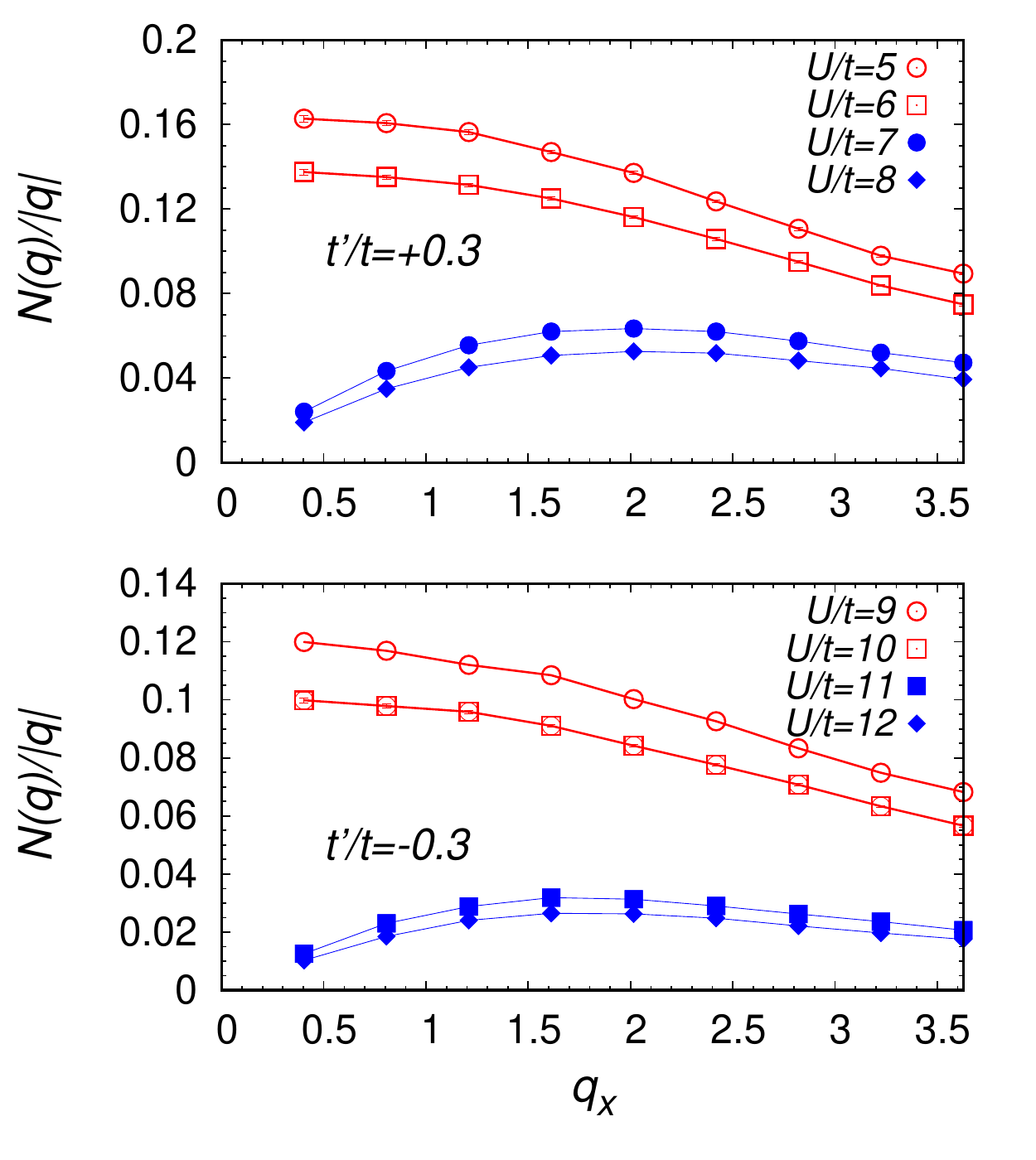}
\caption{Static density-density structure factor $N({\bf q})$, divided by $|{\bf q}|$, over the optimal wave function at different values of 
$U/t$, for $t^\prime/t=+0.3$ (upper panel) and $t^\prime/t=-0.3$ (lower panel). Data are shown for the $L=18\times 18$ lattice size, along 
the line connecting ${\bf \Gamma}=(0,0)$ to ${\bf M}=(\pi,\frac{\pi} {\sqrt{3}})$. Error bars are smaller than the symbol size.}
\label{fig:Nq}
\end{figure}

\section{Results}\label{sec:results}

We first compare the variational energies of different spin-liquid and magnetic phases for $t^\prime/t=+0.3$ and $-0.3$ (corresponding to 
a spin-liquid phase in the Heisenberg model~\cite{zhu2015,hu2015,iqbal2016}). Despite the same large-$U$ limit, the two cases behave in a 
very different way, as shown in Fig.~\ref{fig:energy}. For $t^\prime/t=+0.3$, the spin-liquid regime is confined to the range $U/t \gtrsim 24$, 
while the 120$^{\circ}$ magnetic order is favored for smaller values of $U/t$, down to the Mott transition that occurs at $U_c/t=6.5 \pm 0.5$. 
The location of the Mott transition is determined by looking at the density-density structure factor of Eq.~(\ref{eq:Nq}), see Fig.~\ref{fig:Nq}.
For small values of $U/t$, $N({\bf q})/|{\bf q}|$ extrapolates to a finite value for $|{\bf q}|\to 0$, indicating that the system is metallic; 
instead, for large values of $U/t$, $N({\bf q})/|{\bf q}| \to 0$ for $|{\bf q}|\to 0$, indicating that the system is insulating~\cite{capello2005}. 
By contrast, for $t^\prime/t=-0.3$, the spin-liquid phase extends down to $U/t \approx 16$. Then, for $11 \lesssim U/t \lesssim 16$, the best 
state is the magnetic one with collinear order down to the Mott transition, see Fig.~\ref{fig:Nq}. In both cases, the optimal spin-liquid wave 
function is the one with a nematic $d$-wave symmetry in $\Delta_k$ (see above); instead, the state with a complex pairing has always a higher 
variational energy. Furthermore, in both cases, the magnetic state with collinear order has a lower energy than the spin-liquid one close to 
the Mott transition. This feature resembles the spin-wave result of Ref.~\onlinecite{seki2019}, where by increasing either $J^\prime/J$ or 
$K/J$, the collinear order is favored with respect to the coplanar 120$^{\circ}$ one.

\begin{figure}
\includegraphics[width=0.9\textwidth]{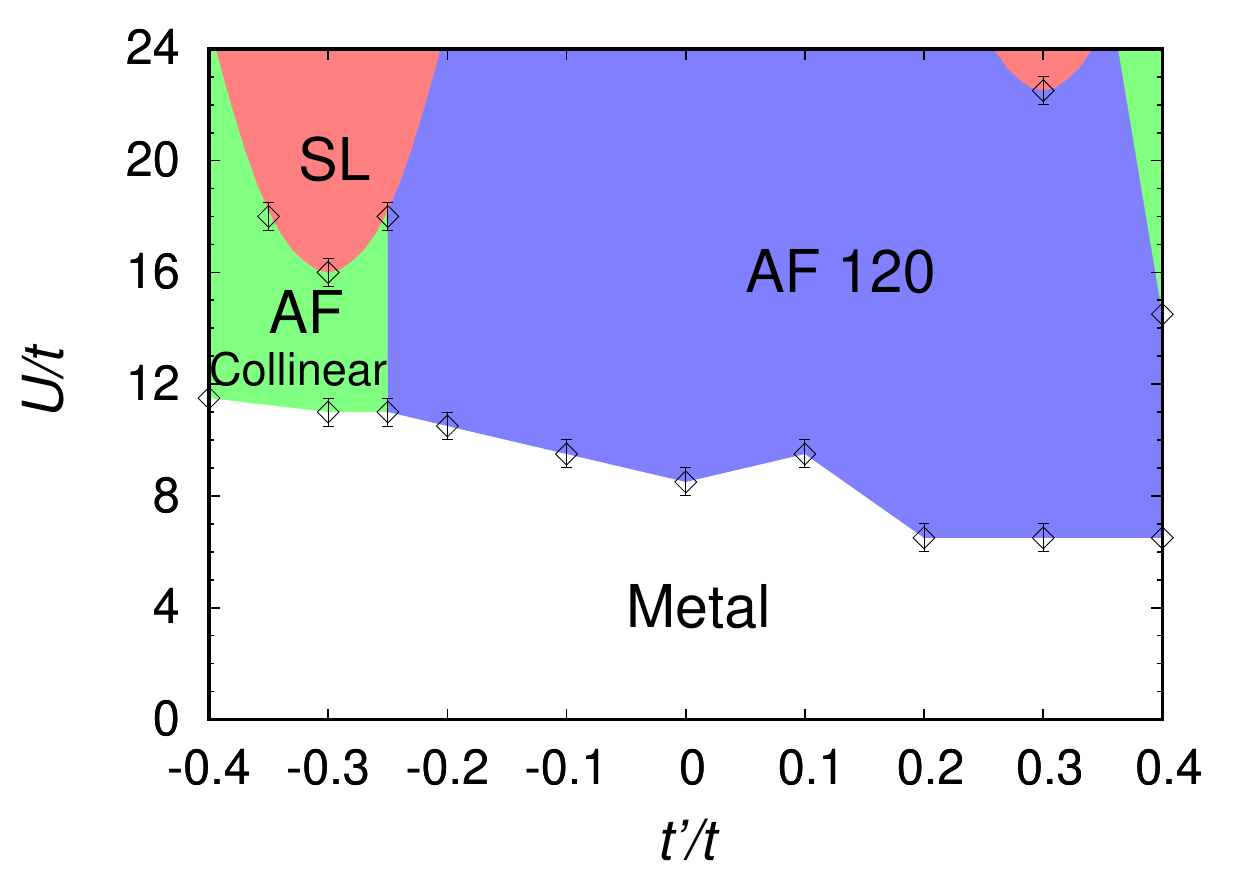}
\caption{Ground-state phase diagram of the $t-t^\prime$ Hubbard model on the triangular lattice at half filling. The magnetic phases are
denoted by blue (for 120$^{\circ}$ order) and green (for stripe collinear order) regions; the spin-liquid phase (with $d$-wave symmetry) is 
denoted by the red region; finally, the white part denotes the metallic phase. Points (with errorbars) indicate the places where phase 
transitions have been located by our calculations.} 
\label{fig:phase}
\end{figure}

\begin{figure}
\includegraphics[width=1.0\textwidth]{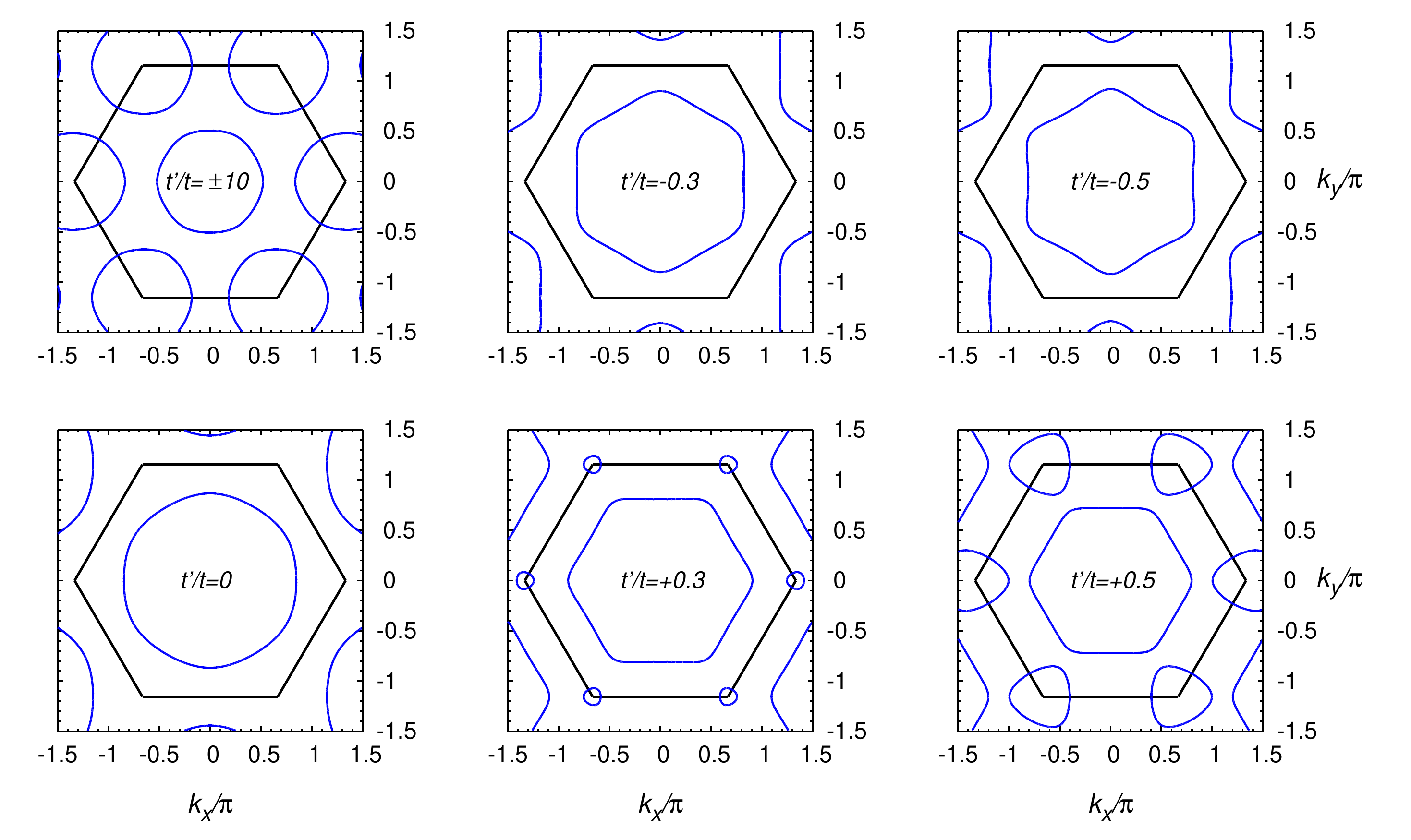}
\caption{Fermi surface at $U=0$, for different values of $t^\prime/t$, in the $(k_x,k_y)$ plane. The first Brillouin zone is denoted by black 
lines, while the Fermi surface is drawn in blue.}
\label{fig:FS}
\end{figure}

In Fig.~\ref{fig:phase}, we report the ground-state phase diagram in the $(t^\prime/t,U/t)$ plane, as obtained by comparing different variational
wave functions. All the phase transitions are first order, since both phases can be stabilized on both sides of the transition. The only 
exception is the one between the metal and the magnetic insulator with 120$^{\circ}$ order that is more compatible with a continuous phase 
transition. In the phase diagram, there is a remarkable asymmetry between the case with positive and negative $t^\prime/t$, which can be 
summarized in these three points: i) the Mott transition is located at smaller values of $U/t$ for $t^\prime/t>0$, ii) the coplanar 120$^{\circ}$ 
order is favored (over the stripe collinear one) for $t^\prime/t>0$, and iii) the spin-liquid phase (with $d$-wave nematic symmetry) is 
stabilized mostly for $t^\prime/t<0$. The first two aspects may be approached from a weak-coupling point of view. In this respect, we report 
in Fig.~\ref{fig:FS} the $U=0$ Fermi surface of the model for different values of the ratios $t^\prime/t$. Starting from an almost circular 
shape at $t^\prime=0$, the Fermi surface evolves in a different way for positive and negative values of $t^\prime/t$. In particular, for 
$t^\prime/t\gtrsim 0.3$, we observe the formation of pockets around the corners of the first Brillouin zone. These pockets are connected by 
vectors that are approximately the ones corresponding to the formation of 120$^{\circ}$ order. The presence of these pockets may lead the 
Mott transition to be located at much lower values of $U/t$ for $t^\prime/t\gtrsim 0.2$ than for smaller values. Note that in the limit of 
$|t^\prime| \gg t$, the Fermi surface is formed by circles around the corners of the first Brillouin zone, corresponding to the limit of a 
triangular lattice defined on NNN bonds with a unit cell that is three times larger than the original one.

\begin{figure}
\includegraphics[width=0.9\textwidth]{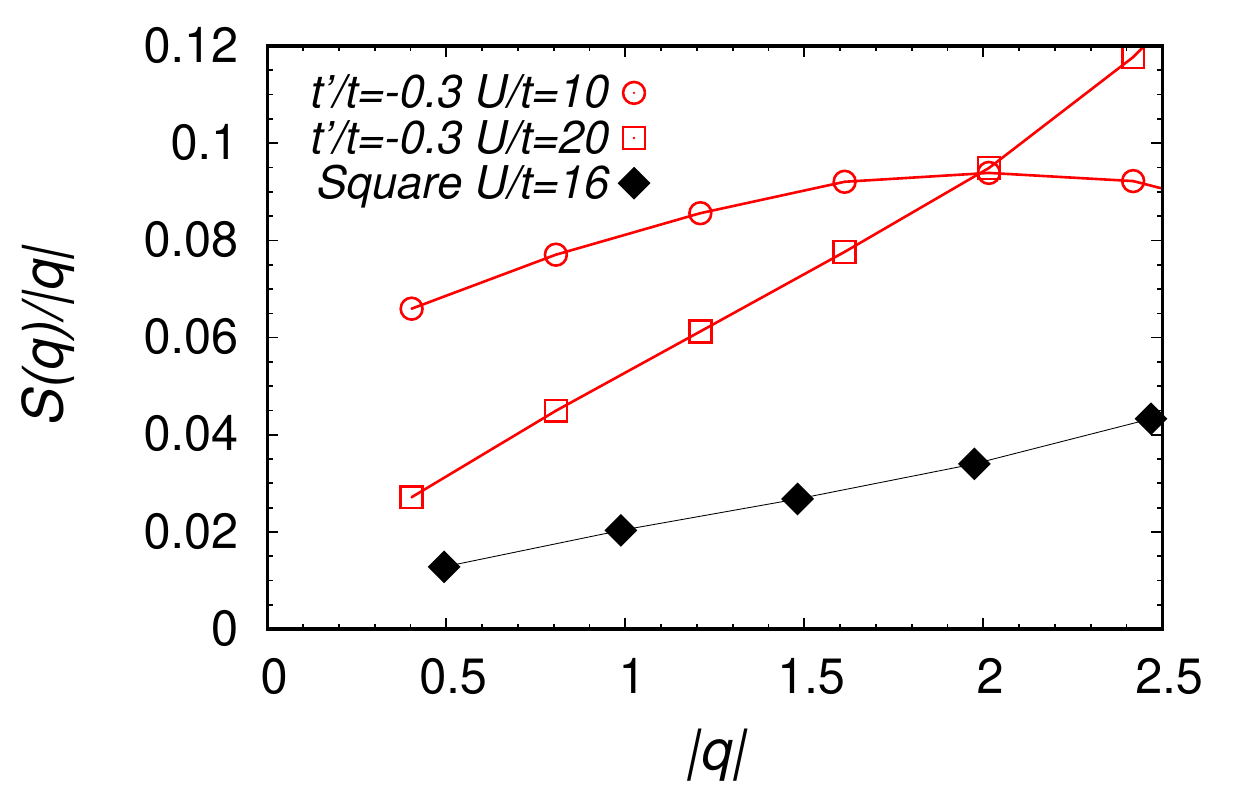}
\caption{Static spin-spin structure factor  $S({\bf q})$, divided by $|{\bf q}|$, over the optimal wave function at $t'/t=-0.3$, $U/t=10$ 
(red empty circles) and at $t'/t=-0.3$, $U/t=20$ (red empty squares), shown along the line connecting ${\bf \Gamma}=(0,0)$ to 
${\bf M}=(\pi,\frac{\pi} {\sqrt{3}})$. $S({\bf q})/|{\bf q}|$ is also shown on the frustrated square lattice at $U/t=16$ from ${\bf \Gamma}=(0,0)$ 
to ${\bf M}=(\pi,\pi)$. All data are presented on a $L=18\times 18$ lattice size. Error bars are smaller than the symbol size.}
\label{fig:Sq}
\end{figure}

Regarding the previous point ii), a clear outcome of our variational approach is that for $t^\prime/t>0$ charge fluctuations favor the 
120$^{\circ}$ magnetic order over the stripe one, as obtained for $t^\prime/t=+0.4$. Here, while for large values of $U/t$ the collinear 
order has the lowest variational energy, for $6.5 \lesssim U/t \lesssim 14.5$ the best wave function is instead the one with coplanar order, 
see Fig.~\ref{fig:phase}. Indeed, also from Fig.~\ref{fig:energy}, which reports the case with a slightly smaller ratio $t^\prime/t=+0.3$, it
is evident that the collinear order is never competitive with the coplanar one, close to the Mott transition. The situation is rather different 
in the opposite side of the phase diagram, where the wave function with collinear magnetic order performs much better and gives the lowest 
variational energy in a wide region. Indeed, for $t^\prime/t \lesssim -0.25$, it can be stabilized down to the metal-insulator transition, 
which takes place for $U_c/t \approx 12$. 
 
Most importantly, a quite large spin-liquid region exists for a sufficiently large electron-electron repulsion and $t^\prime/t<0$ (while it 
is confined to much larger values of $U/t$ for positive ratios of the hopping parameters). We should stress the fact that the nature of this
spin-liquid state is different from the one found by a similar variational approach in the frustrated Heisenberg model~\cite{iqbal2016}. In 
the Hubbard model, hopping and pairing terms break the rotational symmetry, see Eqs.~(\ref{eq:hopp}) and~(\ref{eq:pair}), thus leading to a 
nematic state; this feature is characterized by a convenient order parameter, which can be constructed from the nearest-neighbor spin-spin 
correlations along ``weak'' and ``strong'' bonds, see table~\ref{table:nematic}. Indeed, the bond ${\bf a}_1$, along which pairing and hopping 
in the variational state are suppressed, is characterized by spin-spin correlations that are markedly different from the ones along ${\bf a}_2$ 
and ${\bf a}_3$, along which pairing and hopping are finite. Instead, in the Heisenberg model, the optimal variational wave function contains only 
hopping with a $2 \times 1$ unit cell to accommodate a $\pi$-flux through upward (or downward) triangles. The nematic $d$-wave state can be 
also stabilized, but it has a slightly higher variational energy compared to the best $\pi$-flux {\it Ansatz}. It should be mentioned that the 
latter wave function does not break translational and rotational symmetries {\it only} when limited in the subspace without double occupations 
(suitable for the Heisenberg model). Within the Hubbard model (i.e., in the presence of charge fluctuations), breaking the translational symmetry 
gives rise to a sizable energy loss. Our present results suggest that charge fluctuations will favor the nematic $d$-wave state, thus limiting 
the $\pi$-flux state to exceedingly large values of $U/t$, i.e., much larger than the ones that have been considered here. An aspect that is 
shared between these two spin liquids is the existence of gapless excitations, which can be assessed from the small-$q$ behavior of the spin-spin 
structure factor, see Fig.~\ref{fig:Sq}. Even though the value of $S({\bf q}/|{\bf q}|)$ for $|{\bf q}| \to 0$ shown in the spin-liquid phase 
(at $U/t=20$) is much smaller than the one obtained in the metallic regime (at $U/t=10$), the extrapolation is still compatible with a finite 
value, not much different from the one obtained in the frustrated square lattice, where a gapless spin liquid was found~\cite{tocchio2008}.

\begin{table}
\caption{\label{table:nematic} 
Nearest-neighbor spin-spin correlations $\langle S^z_i S^z_j\rangle$ between sites at positions ${\bf R}_i$ and ${\bf R}_j$, that are connected 
by the nearest-neighbor vectors ${\bf a}_1$, ${\bf a}_2$, and ${\bf a}_3$. Data are computed within the spin-liquid phase at $U/t=20$ and 
$t^\prime/t=-0.3$. Within the error bar, results are the same on four lattice sizes: $6 \times 6$, $10\times 10$, $14\times 14$, and
$18\times 18$}
\begin{tabular}{cc}
\hline
${\bf R}_j$ & $\langle S^z_i S^z_j\rangle$ \\
\hline\hline
${\bf R}_i+{\bf a}_1$  & 0.16(1)   \\
${\bf R}_i+{\bf a}_2$  & -0.40(1)  \\
${\bf R}_i+{\bf a}_3$  & -0.40(1)  \\
\hline \hline 
\end{tabular}
\end{table}

Finally, we would like to mention that metallic, magnetic and spin-liquid wave functions have similar energy variances 
$\sigma_H^2= 1/L (\langle {\cal H}^2 \rangle - \langle {\cal H} \rangle^2)$, this quantity testifying the accuracy of the variational calculation.
In fact, $\sigma_H^2$ is always positive and vanishes only when the variational state is an exact eigestate of the Hamiltonian, e.g., the ground 
state. For example, we find that $\sigma_H^2 \approx 0.1$ in the metal for $t^\prime/t=+0.3$ and $U/t=4$; $\sigma_H^2 \approx 0.1$ in the 
120$^{\circ}$ magnetic phase for $t^\prime/t=+0.3$ and $U/t=20$; $\sigma_H^2 \approx 0.2$ in the spin-liquid regime for $t^\prime/t=-0.3$ and 
$U/t=20$; $\sigma_H^2 \approx 0.1$ in the collinear antiferromagnet for $t^\prime/t=-0.3$ and $U/t=12$. These results suggest that the phase 
diagram should not be much affected by the (slightly) different accuracy of the variational wave functions. Futthermore, the larger variance 
of the spin-liquid state with respect to the other states would indicate that its actual stability region could be broader than what obtained 
in Fig.~\ref{fig:phase}.

\section{Conclusions}\label{sec:concl} 

We have presented the VMC phase diagram of the $t-t^\prime$ Hubbard model on the isotropic triangular lattice, as summarized in Fig.~\ref{fig:phase}, 
which may be relevant for the physics of the transition metal dichalcogenide 1T-TaS$_2$. We found that for $t^\prime/t \approx -0.3$ a spin-liquid 
phase is present down to intermediate values of $U/t$. This phase is nematic and presumably gapless and is not directly connected to the metallic 
state, from which it is separated by a magnetic insulator with collinear order. On the contrary, for positive values of $t^\prime/t$ the coplanar 
magnetic state with 120$^{\circ}$ order dominates the phase diagram. Our calculations do not show any evidence for a weak-Mott insulating phase, 
intruding between the metallic and the antiferromagnetic phases, in contrast with other numerical approaches. We surmise that the high correlation 
of electrons at short/medium distances in the metal close to the metal-insulator transition may lead to the misconceived conclusion of the existence 
of an intermediate spin-liquid phase.

Our results bring a twofold message: On one side the degree of frustration $(t^\prime/t)^2$, already considered in the Heisenberg model, 
drives the appearance of the spin-liquid phases, since no spin liquid is observed for $t^\prime=0$ (e.g., charge fluctuations are not able 
to destroy the magnetic long-range order). On the other side, the sign of $t^\prime/t$, which cannot be detected within the Heisenberg model, 
is crucial to stabilize a spin liquid down to intermediate values of the electron-electron repulsion. In addition, also the nature of the 
magnetically ordered phases (i.e., their periodicity) strongly depends upon the sign of the next-nearest-neighbor hopping.

\acknowledgments

F.B. acknowledges the program ``Topological Quantum Matter: Concepts and Realizations'', held at Kavli Institute for Theoretical Physics (KITP),
partially supported from the National Science Foundation under Grant No. NSF PHY-1748958.

\end{document}